\documentclass[11pt]{article}
\usepackage{setspace}
\usepackage{amssymb}
\usepackage{amsmath}
\usepackage{amsfonts}

\def\be{\begin{equation}}

\def\ee{\end{equation}}

\def\dd{\partial}

\def\bea{\begin{eqnarray}}

\def\eea{\end{eqnarray}}

\newcommand\eps{\epsilon}

\begin{document}

\singlespace

\begin{flushright} BRX TH-669 \\
CALT 68-2945
\end{flushright}

\vspace*{.3in}

\begin{center}

{\Large\bf Stressless Schwarzschild }

{\large S.\ Deser}

{\it 
Lauritsen Laboratory, California Institute of Technology, Pasadena, CA 91125; \\
Physics Department,  Brandeis University, Waltham, MA 02454 \\
{\tt deser@brandeis.edu}
}

\end{center}

\begin{abstract}
This self-contained pedagogical derivation of the Schwarzschild solution, in
``$3 + 1$" formulation and conformal spatial gauge, (almost) avoids all affinity, 
curvature and index gymnastics.
 
  \end{abstract}

\section{Introduction}
The derivation of the Schwarzschild (S) solution in S coordinates has become a quite compact and accessible process (see, eg, [1]). Still, it does involve some tedious calculations before reaching the final, simple, answer. Besides, it is 
always instructive to have alternate paths to a fundamental result: The present one obtains the solution in a different frame, does so with 
a minimum of calculation, notably without the tedious process of computing affinities, curvature components and index-moving, while 
naturally introducing some useful tools, especially the ADM ``$3+1$"  formulation of GR, and (while not strictly necessary), the conformal curvature 
tensor, in its simplest, $D=3$ Cotton-Weyl, incarnation.

In outline, our strategy is to avoid having to deal with the full $4$-metric's components and the attendant messy, if elementary, calculations. The $3+1$ approach is a first step.  To tame the ``$3$", we need a spatial gauge with the simplest possible $3$-metric; conformally flat will fit that bill. Then we will use (in $D=3$) the Einstein tensor's universal properties: identical conservation and homogeneity -- in both (second) derivative, and in $(0)$ metric, orders -- since it depends only on the affinity 
$\Gamma \sim g^{-1} \dd g$ and its derivatives.  These properties will almost completely determine its form. A simple final redefinition of the conformal factor ``linearizes" the $3$-scalar curvature, reducing the relevant equations to (flat) Laplace form, to yield  the usual $m/r$ dependences.

\section{The ADM Action and static $3$-metrics}

The Einstein-Hilbert (EH) action of GR is $4$-covariant, while ADM's [2] is ``$3+1$", hence better suited to intrinsically $3+1$ questions such as ours: 
finding static, that is time-independent and space-time diagonal, $g_{0i}=0$, solutions.  It reads 
\begin{equation}
I_{ADM} [\pi^{ij}, g_{ij}; N, g_{0i}]=\int d^4x \left\{ \pi^{ij} \dot g_{ij} +N\left[ \sqrt{g} R + \left(\pi_{ij}^2-\frac{1}{2} \left(\pi_i^i\right)^2\right)/\sqrt{g} \right]+\frac{1}{2} g_{0i} D_j \pi^{ij}\right\};  
\end{equation}
all $16$ fields are to be varied independently. The six $g_{ij}$ are the spatial components of the covariant metric $g_{\mu \nu}$, while their conjugate momenta $\pi^{ij}$ are given by the relevant field equations in this first order, 
``$p \dot q$", form. All quantities and operations in (1) and henceforth are in the intrinsic $3$-space: no $4D$ objects appear. The static requirement annihilates the $9$ variables $(\pi^{ij} , g_{0i})$;  then $N$, originally defined as $1/\sqrt{-g^{00}}$,  becomes $\sqrt{-g_{00}}$. The action (1) effectively reduces to
\begin{equation}
I_{ADM} (g_{ij}; N) \longrightarrow \int d^3x \sqrt{g} N R(g_{ij}),               
\end{equation}
 and the $16$ Einstein equations immediately shrink to the seven that would follow from (2), 
\begin{subequations}
\begin{align}
 R&=0, \\
(D_i D_j- g_{ij} \nabla^2) N + N G_{ij} &= 0  \, \, \, \, \, \, \, \, \, \, \longrightarrow \nabla^2 N = 0.
\end{align}
\end{subequations}
Obtaining (3b) from varying (1) or (2) mirrors the standard, elementary, derivation of Einstein's equations from the 4D EH action. Both rely on the Palatini identity, valid in any $D$, for the (trace of the) variation of the curvature:
$g^{\mu \nu} \delta R_{\mu\nu} \dot = (g^{\mu \nu} D^2- D^\mu D^\nu) \delta g_{\mu\nu}$. This means that the EH action's curvature variation does not contribute, being a total divergence; only the explicit metric part's ($\sqrt{-g} g^{\mu\nu}$) variation, $\sqrt{-g}  G^{\mu\nu} \delta g_{\mu\nu}$, does.  However in our case, $N$ -- the coefficient of $R$ in (2) ``catches" the curvature's variation upon integration by parts, yielding precisely the first, ``extra", term of (3b). [Incidentally, the Palatini identity itself results immediately just from the definition of $R_{\mu\nu} \sim d \Gamma + \Gamma \Gamma$, and the fact that varying means linearizing, both for curvature and affinity; that is, their respective variations are: 
$\delta R_{\mu \nu} \dot =  D_\alpha \delta \Gamma^\alpha_{\mu \nu} -D_\mu \delta \Gamma _\nu$,  
$\delta \Gamma^\alpha_{\mu \nu}  \dot =  1/2 (- D^\alpha \delta g_{\mu \nu} +..)$. Palatini then follows upon inserting $\delta \Gamma$ into $\delta R$]

We now specialize to spherical symmetric $3$-metrics, whose intervals can be written, in terms of two functions of $r$, as
\begin{subequations} 
\begin{align}
dl^2= A dr^2 + B r^2 d \Omega.
 \end{align}
 Radial gauge choices leave a single unknown;  one of these is the conformal interval
 \begin{align}
  dl^2= e^{2\omega(r)} (dx^2+dy^2+dz^2), \, \, \, \, \, \, \, \, \, \,  g_{ij} = e^{2 \omega} \delta_{ij}.
  \end{align}
 \end{subequations}

[Although it is overkill, we use this occasion to introduce the ($D=3$) conformal curvature, namely the Cotton tensor density $C^{ij}$, defined in elementary texts [3] (see also [2]); both it and its relative, the Schouten tensor $S_{ij}$, are important in numerous other contexts:
\begin{equation}
C^{ij} \dot= \eps^{ilm} D_l S_m^j =C^{ji} \, \, \, \, \, \, \, \, \, \,    S_{mj} \dot= R_{mj}- \frac{1}{4} g_{mj} R.   
\end{equation}
A $3$-metric is conformally flat if and only if its Weyl tensor, here $C^{ij}$, vanishes, which will -- pedantically -- justify the second form in (4b). That $C^{ij} = 0$ for any spherically symmetric metric follows without calculation: 
First note that the generic spherical metric of (4a) may also be written as 
$g_{ij} (r)=\dd^2_{ij} Y(r) + e^{2\omega(r)} \delta_{ij}$. Now consider $C^{ij}$ in Riemann local coordinates, 
where the Ricci and Cotton tensors are (linear) gauge invariant, hence insensitive to the first, pure gauge, part of the metric, leaving just its desired conformal part. QED. (A fancier, but equivalent, derivation would express the metric's gauge part in terms of the radial Killing vectors.)] 

\section{The Solution}
We now solve the field equations (3) for the two unknowns $(\omega,N)$.  The standard derivations of S involve the 
dreaded process of first computing the affinities, then the curvature components etc., of the $4$-metric. We (essentially) avoid these steps:  
As we noted at the outset, our ($3$-space) Einstein tensor $G_{ij}$, is -- as in any $D$ -- identically conserved and homogeneous of second derivative order and of metric order $0$, since it only involves the Christoffel symbols, $\Gamma \sim g^{-1} \dd g$ and their derivatives: These properties almost completely specify $G_{ij}$;  since it depends on (two) derivatives of $\omega$ (but not $\omega$ itself), it must have the general form
\begin{subequations}
\begin{align}
G_{ij} (\omega) = (\dd_i \dd_j -\delta_{ij} \nabla^2) \omega - a \omega_i \omega_j;
\end{align} 
here and below all derivatives, including the Laplacian, are ordinary, and $\omega_i \equiv \omega_{,i}$.  [The only other allowed candidate term, $b \delta_{ij} (\omega_k \omega_l \delta^{kl})$, could have been carried, but would turn out not to be needed.]  The scalar curvature density is  
\begin{align}
 \sqrt{g} R = 2 e^\omega [2 \nabla^2 \omega +  a (\omega_k \omega_l \delta^{kl})]. 
\end{align}
\end{subequations}
Next, we fix $a$ by the Bianchi identity; defining $\mathcal{G}^{ij} \equiv \sqrt{g} G^{ij} = e^{-\omega} G_{ij}$, and using the obvious, all-component,
\begin{equation}
\Gamma^i_{kj} = \delta^i_j \omega_k + \delta^i_k \omega_j - \delta_{kj} \omega_i,
\end{equation}
we get
\begin{equation}
0= D_j \mathcal{G}^{ij} =\dd_j \mathcal{G}^{ij} + \Gamma ^i_{jk} \mathcal{G}^{jk} = (1-a) e^{-\omega} [\omega_{ij} \omega_j +\omega_i \nabla^2 \omega] \, \, \, \, \, \, \, \, \, \longrightarrow a = 1.
\end{equation}
[The covariant divergence of a contravariant vector density is just its ordinary one, so we only needed to face the one $\Gamma$ shown.] 
Having determined the curvature, we now (for convenience) ``linearize" $R$ of (6b) by field-redefining $(\nabla \omega)^2$ away;  writing $e^\omega= \Phi^n$; a $1$-line calculation finds $n=2$:  $g_{ij}= \delta_{ij} \Phi^4$ yields $\sqrt{g} R\sim \Phi \, \nabla^2 \Phi$.

Now we reap the benefits -- immediately solving the field equations. First, (3a):
  \begin{equation}
0= \sqrt{g} R =\Phi \nabla^2 \Phi \longrightarrow \nabla^2 \Phi = 0 \longrightarrow \Phi = c  (1 + m/2 r);
\end{equation}
the constant $c$ is absorbed by spatial coordinate rescaling, and we have written the other constant per the usual convention. [Unlike in S coordinates, where only first derivatives occur, here both $N$ and $\Phi$ obey second order equations, but again only one integration constant will, of course, survive.] Next, since the trace of (3b) says $N$ is covariantly harmonic, or equivalently (flat) $\nabla^2(N \Phi)=0$, we infer
\begin{equation}
[\Phi^2 r^2 N']'=0 \longrightarrow N = b(1+ m'/2r)/(1+m/2r).
\end{equation}
Time-rescaling removes $b$, while the remaining integration constant $m'$ is fixed with little labor by the traceless part of (3b): we simply look at its linearized version, in terms of $\phi \dot= \Phi-1= m/2r$, and $n\dot = N-1=(m'-m)/2r$ by (10).  
Since $\nabla^2 1/r =0$, the remaining, $\dd_i \dd_j$, terms must vanish.  But $n_{ij} + G_{ij} ( \sim 2 \phi_{ij}) = 0$ implies $m' = -m$, so the complete S interval is
\begin{equation}
ds^2 = - [(1-m/2r)/(1+m/2r)]^2 dt^2 + (1+m/2r)^4 (dx^2 + dy^2 + dz^2).
\end{equation}
 
\section{Conclusion}
To summarize, the present exercise entailed less calculation than the usual S derivation, by taking advantage of the ``$3+1$" formulation of GR
and using conformal $3$-frame. It introduced the ``$3+1$" GR approach and (if superfluously) the $D=3$ conformal curvature-Cotton, tensor. Two final, tangential, remarks:  First, we assumed, rather than derived, Birkhoff's theorem (necessary time independence of GR spherical solutions); a proof would be easy enough, if a bit longer. Second, a historical point: why spherically symmetric stress tensors ``have no hair" is rarely -- but should be -- brought up and explained -- spherical static $T_{ij}$ matter sources are not forbidden, so why don't they affect the exterior metric?  The (elementary) answer was spelled out long ago, in [4].

\subsection*{Acknowledgements}
I thank R Beig, and J Franklin for useful discussions. This work was supported by Grants NSF PHY-1266107 and DOE DEFG02-16493ER40701.

\end{document}